\documentclass{mecbic}
\providecommand{\anno}{2011}
\providecommand{\firstpage}{73}
\providecommand{\lastpage}{82}

\usepackage{breakurl}
\usepackage{amsmath,latexsym,verbatim}

\newtheorem{theorem}{Theorem}

\newtheorem{definition}{Definition}

\newtheorem{remark}{Remark}

\newcommand{\ra}{\rightarrow}
\newcommand{\p}{\prime}
\newcommand{\vb}{\mid}
\newcommand{\qed}{\hspace*{1mm}\hfill$\Box$ }

\title{Further Results on Languages of Membrane Structures}
\author{Rama Raghavan
\institute{Department of Mathematics\\Indian Institute of Technology Chennai\\Chennai, India}
\email{ramar@iitm.ac.in}
\and H. Ramesh
\institute{Department of Mathematics\\Indian Institute of Technology Guwahati\\Guwahati, India}
\email{ramesh\_h@iitg.ernet.in}
\and Marian Gheorghe
\institute{Department of Computer Science\\The University of Sheffield\\Sheffield, UK}
\email{M.Gheorghe@dcs.shef.ac.uk}
\and Shankara Narayanan Krishna
\institute{Department of Computer Science and Engineering\\Indian Institute of Technology Bombay\\Powai, Mumbai, India}
\email{krishnas@iitb.ac.in}
}
\def\titlerunning{Further results on Languages of Membrane Structures}
\def\authorrunning{R. Raghavan, H. Ramesh, M. Gheorghe and S.N. Krishna}

\begin{document}
\maketitle
\pagestyle{plain}
\pagenumbering{arabic}
\setcounter{page}{73}

\begin{abstract}
In \cite{lang-AM}, P systems with active membranes
were used to generate languages, in the sense of languages associated with the
structure of membrane systems. Here, we analyze the power of P systems with
membrane creation and dissolution restricted to elementary membranes, P systems
without membrane dissolution operating according to certain output modes. This leads us to
characterizations of recursively enumerable languages.
\end{abstract}

\section{Introduction}
\par In \cite{lang-AM}, an alternative approach to generate languages by means
of P systems was proposed. An appropriate representation for a string was built
by means of a membrane structure and then the string is generated by visiting the
membrane structure according to a well-specified strategy. P systems with active
membranes were considered, allowing membrane creation or division or duplication
and dissolution, where the output of a computation may be obtained either by
visiting the tree associated with the membrane structure, or by following the
traces of a specific object, called traveller, or sending out the objects. For
each of these approaches, characterizations of recursively enumerable languages
were provided based on P systems that use different sets of operations for
modifying the membrane structure.
\par The output of a computation was considered not as a single entity, which is
either sent out of the system or collected in a specific membrane. Instead the
output is given by catenating the content or the labels of each region of the
whole configuration reached by the system at the end of a computation.  They
considered a general class of P systems with active membranes equipped with
membrane division, creation, duplication and dissolution operations. Membrane
duplication means that, starting from an existing membrane, we can create a new
membrane which encloses the existing one. Then three different approaches for
collecting the output of a computation was given namely visiting the tree
associated with the membrane structure, following the traces of a special object
(traveller traces), and sending out the objects (external mode). The trace mode
and external mode were investigated earlier in the literature. For the external
mode, the main difference with respect to this approach is that, before sending
out the objects, we need to prepare an appropriate membrane structure where the
output objects are supposed to be distributed according to a specific strategy.
\par The approach presented in \cite{lang-AM}, is related to the problem of
finding alternative ways to define the output of the computation in membrane
systems. In fact, this method puts emphasis on the structure of the membranes whose
role is important in successful computations.
\par In this paper, we investigate the computational power of P systems with active
membranes equipped with membrane creation and membrane dissolution restricted to
elementary membranes operating according to all four output modes.  Also we analyse
the power of P systems without membrane dissolution operating according to the three
identified output modes.  We need the label changing feature of {\it in} type rules to
obtain the universality in the second case.
\par The paper is organized as follows.  Section 3 recalls the definition of P
systems
with active membranes together with the definition of three different output
modes.  In section
3.1 we state the results from \cite{lang-AM} concerning the power of P systems
with active membranes
generating languages of membrane structures. In sections 4 and 5, we prove
characterizations
of recursively enumerable languages by means of P systems with active membranes
equipped with
the membrane creation and dissolution operations.
\section {Some Prerequisites}
In this section we introduce some formal language theory notions
which will be used in this paper; for further details, refer to
\cite{handbook}.
\par
For an alphabet $V$, we denote by $V^*$ the set of all strings
over $V$, including the empty one, denoted by $\lambda$.  By $RE$ we
denote the family of recursively enumerable languages.
\par
In our proofs in the following sections we need the notion of a
{\em matrix grammar with appearance checking}. Such a grammar is a
construct $G = (N, T, S, M, F)$, where $N, T$ are disjoint
alphabets, $S \in N$, $M$ is a finite set of sequences of the form
$(A_1 \ra x_1, \ldots, A_n \ra x_n), n \geq 1$, of context free
rules over $N \cup T$, and $F$ is a set of occurrences of rules in
$M$ ( $N$ is the nonterminal alphabet, $T$ is the terminal
alphabet, $S$ is the axiom, while the elements of $M$ are called
matrices).
\par
For $w, z \in (N \cup T)^*$ we write $w \Rightarrow z$ if there is
a matrix $(A_1 \ra x_1, \ldots, A_n \ra x_n)$ in $M$ and the
strings $w_i \in (N \cup T)^*, 1 \leq i \leq n+1$, are such that $w =
w_1, z = w_{n+1}$, and, for all $1 \leq i \leq n$, either (1) $w_i
= w_i^{'} A w_i^{''}$, $w_{i+1} = w_i^{'} x_i w_i^{''}$, for some
$w_i^{'}, w_i^{''} \in (N \cup T)^*$, or (2) $w_i = w_{i+1}, A_i$
does not appear in $w_i$, and the rule $A_i \ra x_i$ appears in
$F$. (The rules of a matrix are applied in order, possibly
skipping the rules in $F$ if they cannot be applied - one says
that these rules are applied in the {\em appearance checking
mode}).
\par
The language generated by $G$ is defined by $L(G) = \{w | w \in T^*,
S \Rightarrow^* w\}$.  The family of languages of this form is
denoted by $MAT_{ac}$.  It is known that $MAT_{ac} = RE$.
\par We say that a matrix grammar with
appearance checking $G = (N, T, S, M, F)$ is in the {\em Z-binary
normal form} if $N = N_1 \cup N_2 \cup \{S, Z, \#\}$, with these
sets mutually disjoint, the matrices of type 3 can also be of the
form $(X \ra Z, A \ra \#)$, and the only matrix of type 4 (terminal
matrix) is of the form $(Z \ra \lambda)$.
\par According to Lemma 1.3.7 in \cite{rrwbook}, for each matrix
grammar there is an equivalent matrix grammar in the binary normal
form.

\par Next we define a computing device which is equivalent in power with Turing
machine.
Such a machine runs a program consisting of numbered instructions of
several simple types. Several variants of register machines with
different number of registers and different instruction sets were
shown to be computationally universal (e.g., see~\cite{minsky}).

\par An \textit{$n$-register machine} is a construct $M=\left(
n,H,l_0,l_h,I\right), $ where:
\begin{itemize}
\item  $n$ is the number of registers,
\item $H$ is the set of instruction labels,
\item  $l_0$ is the initial label,
\item  $l_h$ is the final label, and
\item  $I$ is a set of labelled instructions of the form $l_i:\left( op\left(
r\right), l_j, l_k\right) $, where $op\left( r\right) $ is an
operation on register $r$ of $M$, $l_i,l_j,l_k$ are labels from
the set $H$ (which labels the instructions in a one-to-one
manner),
\end{itemize}

\noindent The machine is capable of the following instructions:
\begin{itemize}
\item[]  $(ADD(r), l_j, l_k)$:  Add one to the contents of register $r$ and
proceed to
instruction $l_j$ or to instruction $l_k;$ in the deterministic
variants usually considered in the literature we demand $l_j=l_k$.
\item[]  $(SUB(r), l_j, l_k)$:  If register $r$ is not empty, then subtract one
from its
content and go to instruction $l_j$, otherwise proceed to
instruction $l_k$.
\item[] {\it halt}:  Stop the machine. This additional instruction can only be
assigned to the final label $l_h$.
\end{itemize}

\par When considering the generation of languages, we use the model of a {\it
register machine with output tape} (e.g., see \cite{artiom}) , which also uses a tape operation:
\begin{itemize}
 \item $l:(write(a), l^{\p\p})$: Write symbol $a$ on the the output tape and go to $l^{\p\p}$.
\end{itemize}

\par We then also specify the output alphabet $T$ in the description of the
register machine with output tape, i.e., we write $M=\left(
n, T, H, l_0, l_h, I\right)$.  Let $L \subseteq V^*$ be a recursively enumerable
language.  Then $L$ can be generated by a register machine with output tape and
with 2 registers.

\section{Languages of Membrane Structures}
We consider a general class of P systems with active membranes equipped with
membrane division, creation, duplication and dissolution operation.  These operations
represent abstractions of cellular biology processes of mitosis and membrane formation
through self-assembling lipid bilayers \cite{cell}. We recall the
definition of this P system from \cite{lang-AM}.
\begin{definition}\label{defn1}
A {\it P system with active membranes} is a construct $$\Pi = (V, K \cup \{0\},
\mu, w_0, w_1, \ldots, w_{m-1}, R)$$ where
\begin{enumerate}
 \item $V$ is an alphabet; its elements are called objects;
 \item $K$ is an alphabet; its elements are called labels; the symbol $0 \notin
K$ is the label of the skin membrane;
\item $\mu$ is a membrane structure containing $m \geq 1$ membranes; the skin
membrane is labelled by 0 and all other membranes are labelled with symbols in
$K$;
\item $w_0$ is the multiset associated with the skin membrane;
\item $w_i \in V^*$, for $1 \leq i \leq m-1$, is a multiset of objects
associated with the membrane $i$;
\item $R$ is a finite set of rules of the form:
\begin{itemize}
 \item[a)] $[_i a \ra v]_i$ with $a \in V, v \in V^*$, and $i \in K \cup \{0\}$
(inside a membrane $i$ an object $a$ is replaced by a multiset $v$),
\item[b)] $[_ia]_i \ra b[_i]_i$ with $a, b \in V$, and $i \in K \cup \{0\}$ (an
object is sent out from a membrane, maybe modified),
\item[c)] $a[_i]_i \ra [_ib]_i$ with $a, b \in V$, and $i \in K \cup \{0\}$ (an
object is moved into a membrane, the object may be modified),
\item[d)] $[_ia \ra [_jb]_j]_i$ with $a, b \in V$, $i \in K \cup \{0\}$, and $j
\in K$ ({\it membrane creation:} inside a membrane $i$, starting from an object
$a$, a new elementary membrane $j$ is created, which contains an object $b$),
\item[e)] $[_ia]_i \ra [_kb]_k[_jc]_j$ with $a, b, c \in V$, $i, j, k \in K$
({\it membrane division:} the membrane $i$, in the presence of an object $a$, is
divided into two new membranes labelled by $k$ and $j$, and the content (objects
and sub-membranes) of the membrane $i$ is copied into each new membrane where
the object $a$ is respectively replaced by $b$ or $c$),
\item[f)] $[_ia]_i \ra [_kb[_jc]_j]_k$ with $a, b \in V$, $i, j, k\in K$ ({\it
membrane duplication} the membrane $i$, in the presence of an object $a$, is
duplicated, that is, the label $i$ is changed into $j$,  the object $a$ is
replaced by $c$, and a new upper membrane labelled by $k$ is created, which
contains an object $b$),
\item[g)] $[_ia]_i \ra a$ with $a \in V$, and $i \in K$ ({\it membrane
dissolution:} in the presence of an object $a$, the membrane $i$ is dissolved and
its content (objects and sub-membranes) is released in the directly upper
region).
\end{itemize}
\end{enumerate}
\end{definition}
\par In the above system we have: an initial membrane structure with $m$
membranes that contain $m$ multisets associated with the regions, and a finite
set of evolution rules. Moreover, as usual in P systems with active membranes,
we also consider a distinct alphabet $K$ which is used to label the membranes
and is necessary to precisely identify the rules that can be applied inside
every membrane. In general, in a P system with active membranes, the number of
membranes can be increased and decreased arbitrarily and there can be many
different membranes with the same label, which can be distinguished from each
other only by the objects they contain. Thus, the labels from $K$ make possible
to keep finite the representation of a P systems by specifying a set of "types",
each one with its own set of rules, for the membrane possibly present in the
system at any time.  A membrane with no further membrane inside is called an
{\it elementary membrane}.
\par The set $R$ contains rules for modifying both the number and the
distribution of objects
inside the system and the number and type of membranes which define the
structure of
the system. The former rules are expressed in the form of transformation and
communication
rules (rules of type (a), (b) and (c)) whereas the latter ones (rules of type
(d),
(e), (f) and (g)) comprise the operations of: membrane creation, membrane
division,
membrane duplication and membrane dissolution, respectively.
\begin{remark}
Here, we do not consider the feature of membrane polarization for P systems with
active membranes as reported in the literature.  However, in the above definition,
rules of more general forms are used that are able to change the labels
of the membranes involved.
\end{remark}
\par As usual, P systems with active membranes evolve according to a
non-deterministic
maximal parallel strategy. Rules of type (a), (b), (c), and (d) are applied to
all the
objects which they can be applied to. Rules of type (e), (f), (g) are applied to
all the membranes which they can be applied to. Obviously, in each step, the
same membrane cannot be used by more than one rule of type (e), (f), (g) (i.e.,
a membrane cannot simultaneously be divided, duplicated and dissolved). More
precisely, we assume that, in each step, the objects first evolve by means of
rules of type (a), (b), (c), (d), and then the membranes evolve according to
rules of type (e), (f), (g).
\par A computation is obtained by applying rules of $R$ starting from the initial
configuration. A computation is said successful if it reaches a configuration where no more
rules can be applied.
\par We illustrate the application of rules (d) - (g) by examples.  In the following examples,
$P, Q$ are possible contents of membranes and $a,b,c$ are objects from $V$. The effect of
the rules on some membrane structures is below:
\begin{itemize}
 \item[(1)] $[_iP~a]_i$\\
 {\it rule of type (d)} : We get $[_iP~[_jb]_j]$\\
 {\it rule of type (e)} : We get $[_jP~b]_j~[_kP~c]_k$\\
 {\it rule of type (f)} : We get $[_kb~[_jP~c]_j]_k$
 \item [(2)] $[_j[_iP~a]_i~Q]_j$\\
 {\it rule of type (g)} : We get $[_j PQ~a]_j$
\end{itemize}

\par The result of a computation may be considered in various forms,
which are called output modes.
\begin{itemize}
 \item {\it Visiting the tree.} The result of a computation is the set of
strings obtained by visiting the tree associated with the membrane structure in
the final configuration. The resulting set of strings is obtained by
concatenating either the labels of the membranes or the objects inside these
membranes, in the order they are visited. If a membrane contains more than one
object, then we consider all the possible permutations of these objects. When we
collect the labels, we do not consider the skin membrane, which is always
labelled by 0. This output mode is denoted either by {\it lab}, if we collect
the labels, or by {\it obj} if we collect the objects.
\item {\it Traveller traces.} We assume that the initial configuration contains
a special object $t$, called the {\it traveller}, inside some membrane. The
traveller $t$ can be moved by using rules of type (b) or (c), but it cannot be
modified by any rule. The resulting string
is obtained as follows: initially we start with the empty string associated with
the
initial configuration, then whenever the object $t$ crosses a membrane labelled
by $i$,
we add the symbol $i$ at the rightmost side of the current string. This output
mode
is denoted by {\it traces}.
\item {\it External mode.} The resulting set of strings is defined as follows:
we start initially with an empty string outside of the membrane system; whenever
an object is sent
out of the skin membrane, we add such an object to the rightmost end of each
current string. If some objects are sent out from the skin membrane at the same
time, we consider the string formed by all the permutations of these objects.
This
output mode is denoted by {\it ext}.
\end{itemize}
\par We denote by $LOP_{m,n}(Op, l)$, with $Op \subseteq \{a, b, c, d, e, f,
g\}$, $l \in \{obj, lab, traces, ext\}$, the family of languages generated by P
systems with active membranes with at most $m$ membranes in the initial
configuration that use at most $n$ different labels for the membranes (the
cardinality of $K \cup \{0\}$ is at most~$n$), that apply rules of the forms
specified in $Op$, and has the output mode $l$. As usual, if the value of $m$,
or the value of $n$, is not bounded, it is replaced by the symbol $*$. Moreover,
when the rules of type $(e), (f)$, and $(g)$ are allowed only for elementary
membranes, the corresponding operations are denoted by $e^\p, f^\p$, and $g^\p$.
Also, when the rules of type $(b)$ and $(c)$ are allowed to change the label of
the membrane, the corresponding operations are denoted by $b^\p$ and $c^\p$.

\subsection{The power of membrane creation and membrane division}
\par In this section, we present some results from \cite{lang-AM}. The universality
of P systems with membrane division and membrane dissolution with respect to the output
modes {\it lab, obj} was given by the following theorem.
\begin{theorem}
 $LOP_{1,*}(\{a, b, c, d, g\}, v) = RE, \mbox{ for } v \in  \{lab, obj\}$.
\end{theorem}
\par A similar result holds for  P systems with membrane division and membrane
dissolution restricted to elementary membranes.
\begin{theorem}
$LOP_{2,*}(\{a, b, c, e, g^\p\}, v) = RE, \mbox{ for } v \in  \{lab, obj\}$.
\end{theorem}

In the proofs of the above theroem, the final configuration has membrane
structure of depth 2.  So a ``predefined'' order among the membranes is necessary
to get a suitable representation for the strings of a language.  Such an approach
does not work well in the case of {\it traces} and {\it external} output modes.  The
following theorem shows that such a problem can be avoided by considering the operation
of {\it membrane duplication}.
\begin{theorem}
 $LOP_{2,*}(\{a, b, c, f, g\}, v) = RE, \mbox{ for } v \in  \{lab, obj, ext\}$
and \\
$LOP_{3,*}(\{a, b, c, f, g\}, traces) = RE$.
\end{theorem}

The following cases were left open in \cite{lang-AM}:\\
The computational power of
\begin{enumerate}
 \item P systems with membrane creation and membrane dissolution (or membrane
division and
membrane dissolution) operating according to the external mode or the traces
mode;
\item P systems without membrane dissolution operating according to any of the
three identified output modes;
\item P systems with membrane creation and membrane dissolution restricted
to elementary membranes.
\end{enumerate}

We settle some of the above cases in the coming sections.

\section{Universality with Membrane Creation and Dissolution}
The following is a characterization of recursively enumerable languages by means of P systems with
active membranes equipped with the membrane creation and dissolution
operation restricted to elementary membranes operating according to all output modes.

\begin{theorem}
 $LOP_{1,*}(\{a, b, c, d, g^\p\}, v) = RE, \mbox{ for } v \in  \{lab, obj,
ext\}$ and \\
$LOP_{2,*}(\{a, b, c, d, g^\p\}, traces) = RE$.
\end{theorem}
{\bf Proof:}  The proof is based on the simulation of a register machine $M =(2,
T, P, l_0, l_h)$.  We construct a P system with active membranes that simulates
the register machine $M$ such that $$\Pi=(V, K \cup \{0\}, [_0]_0, l^\p_0, R)$$
where
\begin{eqnarray*}
V & = & T \cup \{a_1, a_2, b_1, b_2\} \cup \{l^\p, l^{\p\p} \vb l : (ADD(r),
l^\p, l^{\p\p}) \in P\}\\
& \cup & \{l_0, l^\p_0, l^{\p\p}_0, 1^\p, 2^\p, \$^\p \vb  l_0 \mbox{ is the
initial label of M }\}\\
& \cup & \{l_i, l^\p_i, l^{\p\p}_i, l^{\p\p\p}_i, l^{iv}_i, l^\p, l^{\p\p} \vb l
: (SUB(i), l^\p, l^{\p\p}, i=1,2\}\\
& \cup & \{(a,l^\p),\overline{(a,l^\p)} \vb l:(WRITE(a), l^\p) \in P, a \in
T\}\\
& \cup & \{1^\p, 2^\p, \$^\p\}\\
K & = & T \cup \{1, 2, 3, 4, \$ \}
\end{eqnarray*}
\begin{eqnarray*}
R & = & \{[_0l^\p_0 \ra 1^\p 2^\p \$^\p l^{\p\p}_0]_0, [_0l^{\p\p}_0 \ra l_0]
\vb l_0 \mbox{ is the initial label of M} \}\\
& \cup & \{[_0 1^\p \ra [_1]_1]_0, [_0 2^\p \ra [_2]_2]_0, [_0 \$^\p \ra
[_\$]_\$]_0\}\\
& \cup & \{[_0l \ra b_il^\p]_0, [_0l \ra b_il^{\p\p}]_0 \vb l:(ADD(i), l^\p,
l^{\p\p}), i=1,2\}\\
& \cup & \{b_i[_i]_i \ra [_ia_i]_i \vb i=1,2\}\\
& \cup & \{l \ra l_i, l_i[_i]_i \ra [_il_i]_i \vb l:(SUB(i), l^\p, l^{\p\p},
i=1,2\}\\
& \cup & \{a_i[_3]_3 \ra [_3a_i]_3, [_3l^\p_i \ra l^{\p\p}]_3 \vb l:(SUB(i),
l^\p, l^{\p\p}, i=1,2\}\\
& \cup & \{[_3a_i]_3 \ra \lambda, [_3 l^{\p\p}_i \ra l^{\p\p\p}_i]_3,
[_3l^{\p\p\p}_i]_3 \ra l^{iv}_i \vb l:(SUB(i), l^\p, l^{\p\p}, i=1,2\}\\
& \cup & \{[_il^{\p\p\p}_i]_i \ra l^{\p}, [_il^{iv}_i]_i \ra l^{\p\p}\vb
l:(SUB(i), l^\p, l^{\p\p}, i=1,2\}\\
& \cup & \{ [_0l \ra (a, l^\p)]_0 \vb l:(WRITE(a),l^\p)\}\\
& \cup & \{(a,l^\p)[_b]_b \ra [_b(a,l^\p)]_b \vb b \in T \cup \{\$\}\}\\
& \cup & \{ [_\$(a, l^\p)]_\$ \ra \overline{(a,l^\p)}, [_b\overline{(a,l^\p)}
\ra [_al^\p\$]_a]_b \vb b \in T\}\\
& \cup & \{[_a \$ \ra [_\$]_\$]_a \vb a \in T\}\\
& \cup & \{l_h[_1]_1 \ra [_1l_h]_1, [_1 l_h]_1 \ra l_h^\p, l_h^\p[_2]_2 \ra
[_2l_h^\p]_2, [_2 l_h^\p]_2 \ra l_h^{\p\p}\}\\
& \cup & \{[_0a_i \ra \lambda]_0 \vb i = 1,2\}\\
& \cup & \{l_h^{\p\p}[_4]_4 \ra [_4l_h^{\p\p}]_4, [_4 l_h^{\p\p}]_4 \ra
\lambda\}
\end{eqnarray*}
Let us see how the P system $\Pi$ works. Initially, we have the configuration
$[_0[_4t]_4l_0^\p]_0$.  We apply the first 5 rules to produce the configuration
$[_0l_0[_1]_1[_2]_2[_\$]_\$]_0$.  The value of the two registers $i=1,2$, are
represented by the number of objects $a_i$ inside the corresponding membrane
$i$.  The membrane labelled~$\$$ is used to prepare an appropriate membrane
structure where the output objects are supposed to be distributed according to a
specific strategy.
\par The add instruction $l:(ADD(i), l^\p, l^{\p\p})$ is simulated as follows.
We use the rule $l \ra b_il^\p$ or $l \ra b_il^{\p\p}$ to create an object $b_i$
corresponding to the register $i$.  Now the object $b_i$ changes to $a_i$ while
entering inside membrane $i$.
\par In order to simulate a subtract instruction $l:(SUB(i), l^\p, l^{\p\p})$,
we send the object $l_i$ into the membrane~$i$ and then proceed in the following
way: The object $l_i$ creates a membrane with label 3 and an object $l_i^\p$.
If the register $i$ is not empty, then the object $a_i$ will enter membrane 3
and dissolve it; otherwise the object $l_i^{\p\p\p}$ dissolves membrane 3 there
by changing to $l_i^{iv}$.  If the register $i$ is not empty, then we have
$l_i^{\p\p\p}$ in membrane $i$; otherwise $l_i^{iv}$.  Now we will send $l^\p$
or $l^{\p\p}$ to the skin membrane depending upon the presence of $l_i^{\p\p\p}$
or $l_i^{iv}$ in membrane $i$ respectively.  This will end the simulation of the
SUB instruction.
\par The simulation of the instruction $l:(WRITE(a), l^\p)$ is done as follows.
First we use the rule $l \ra (a,l^\p)$ in the skin membrane.  The object
$(a,l^\p)$ travels deep inside the nested membrane structure until it reaches
the membrane $[_\$]_\$$.  In membrane $\$$, the object $(a,l^\p)$ changes to
$\overline{(a,l^\p)}$ and dissolves the membrane.  Now the object
$\overline{(a,l^\p)}$ will create a membrane labelled $a$ which contains the
objects $l^\p$ and $\$$.  The object $l^\p$ moves toward the skin membrane
whereas the object $\$$ will create a membrane $[_\$]_\$$ inside the membrane
$[_a]_a$.  The object $l^\p$ starts the simulation of the instruction labelled
$l^\p$ after reaching the skin membrane.
\par The presence of object $l_h$ in the skin membrane will start the clean-up
process.  It will remove both the membranes 1, 2 and the objects inside them.
Finally the object $l_h^{\p\p}$ dissolves membrane 4 which contains the object
$t$.
\par At last, we have a configuration of the form
$[_0t[_{x_1}[_{x_2}\ldots[_{x_h}[_\$]_\$]_{x_h}\ldots]_{x_2}]_{x_1}]_0 $ with
$x_1x_2\ldots x_h \in L(M)$, for some $h \geq 1$.  Now we move the traveller $t$
by using rules of the form $t[_a]_a \ra [_at]_a$, with $a\in T$, and in this way
we generates exactly the string $x_1x_2\ldots x_h \in L(M)$.
\par \noindent{\it External mode:}  For this mode we consider a P systems $\Pi$
whose initial configuration is $[_0l_0^\p]_0$, where $l_0$ is the starting label
of the register machine $M$.  We simulate the register machine $M$ in the same
way as described above for the traveller traces, and during the clean-up process
the object $l_h^\p$ changes to $f$ and dissolves membrane 2 instead of changing
to $l_h^{\p\p}$.  Thus, we have a configuration
$[_0f[_{x_1}[_{x_2}\ldots[_{x_h}[_\$]_\$]_{x_h}\ldots]_{x_2}]_{x_1}]_0$, and we
can generate the string $x_1x_2\ldots x_h \in L(M)$ by using the following
rules:
\begin{itemize}
 \item $f[_a]_a \ra [_af^\p]_a$ with $a \in T$
 \item $[_a f^\p \ra af]_a$ with $a \in T$
 \item $[_ba]_b \ra a[_b]_b$ with $a,b \in T$
 \item $[_0a]_0 \ra a[_0]_0$ with $a \in T$
\end{itemize}
By applying these rules, we can send the objects out of the skin membrane in the
right order
\par We can easily modify the above system $\Pi$ to obtain a final configuration
of the form\newline
$[_0[_{x_1}x_1[_{x_2}x_2\ldots[_{x_h}x_h[_\$]_\$]_{x_h}\ldots]_{x_2}]_{x_1}]_0$
for some $h \geq 1$, and $x_1x_2\ldots x_h \in L(M)$.  If we visit the tree
associated with this membrane structure either by collecting the labels or by
collecting the objects, then we get $x_1x_2\ldots x_h \in L(M)$ in both
cases.\qed
\begin{remark}
The universality of P systems with membrane division and
membrane dissolution restricted to elementary membranes with respect to
the {\it traces} and {\it external} output modes can be proved in a similar fashion provided
the rules of type {\it endocytosis} were allowed.  Because a combination of
rules of type {\it division} and {\it endocytosis} can simulate rules of type {\it creation}.
\end{remark}

\section{Universality with only Membrane Creation}
\par A similar result holds for P systems that use only the membrane creation
operation avoiding the operation membrane dissolution for all output modes
except {\it traveller traces}.  But we need the label changing feature for {\it
in} type rules to obtain universality.
\begin{theorem}
 $LOP_{1,*}(\{a, b, c^\p, d\}, v) = RE, \mbox{ for } v \in  \{lab, obj, ext\}$.
\end{theorem}
{\bf Proof:}  Let $G = (N, T, S, M, F)$, with $N = N_1 \cup N_2  \cup \{S, Z, \#
\}$, be a matrix grammar with appearance checking in Z-binary normal form where
the matrix of type 1 is $(S \ra XA)$, the matrices of type 2 are labelled, in
one to one manner, by $m_1, \ldots, m_k$, and matrices of type 3 by $m_{k+1},
\ldots, m_n$. We construct a P system with active membranes that simulates the
matrix grammar $G$ as follows:
$$\Pi = (V, K \cup \{0\}, [_0]_0, S, R)$$ where
\begin{eqnarray*}
 V & = & N_1 \cup N_2 \cup T \cup \{Z, \#\} \cup \{Y_i, Y^\p_i\vb 1 \leq i \leq
n, Y \in N_1\}\\
& \cup & \{Y_{i,B} \vb Y \in N_1, B \in N_2, 1 \leq i \leq k \} \cup \{Y_{i,\$}
\vb Y \in N_1, 1 \leq i \leq k\}\\
K & = & N_2 \cup T \cup \{\$\}
\end{eqnarray*}
\begin{eqnarray*}
R & = & \{ [_0 S \ra [_AX_A]_A]_0, [_A X_A \ra [_\$ X^\p_A]_\$]_A \vb (S \ra XA)
\in M\}\\
& \cup & \{[_\$X^\p_A]_\$ \ra X^\p_A[_\$]_\$, [_A X^\p_A]_A \ra X [_A]_A \vb (S
\ra XA) \in M\}\\
& \cup & \{[_0 X \ra Y_i]_0 \vb X, Y \in N_1, 1 \leq i \leq n\}\\
& \cup & \{Y_i[_y]_y \ra [_yY_i]_y \vb y \in N_2 \cup T, y \neq A, m_i : (X \ra
Y, A \ra x) \in M,   1 \leq i \leq k\}\\
& \cup & \{Y_i[_\$]_\$ \ra [_\#\#]_\# \vb Y \in N_1, 1 \leq i \leq k\}\\
& \cup & \{Y_i[_A]_A \ra [_aY^\p_i]_a \vb m_i : (X \ra Y, A \ra a), 1 \leq i
\leq k\}\\
& \cup & \{[_yY^\p_i]_y \ra Y^\p_i[_y]_y \vb y \in N_2 \cup T \cup \{\$\}, 1
\leq i \leq n\}\\
& \cup & \{[_0Y^\p_i \ra Y]_0 \vb Y \in N_1, 1 \leq i \leq n\}\\
& \cup & \{Y_i[_A]_A \ra [_{a_1}Y_{i,a_2}]_{a_1} \vb m_i:(X \ra Y, A \ra
a_1a_2), 1 \leq i \leq k\}\\
& \cup & \{Y_{i,B}[_C]_C \ra [_BY_{i,C}]_B \vb B, C \in N_2 \cup T \cup \{\$\},
1 \leq i \leq k\}\\
& \cup & \{[_B Y_{i, \$} \ra [_\$Y^\p_i]_\$]_B \vb Y \in N_1, B \in N_2, 1 \leq
i \leq k\}\\
& \cup & \{Y_i[_A]_A \ra [_\#\#]_\#, Y_i[_\$]_\$ \ra [_\$Y^\p_i]_\$ \vb m_i:(X
\ra Y, A \ra \#), k+1 \leq i \leq n\}\\
& \cup & \{[_\#\# \ra \#]_\#, [_0 Z \ra \lambda]_0\}
\end{eqnarray*}
\par Initially, we have the configuration $[_0S]_0$.  We simulate the unique
matrix of type 1 in $G$ by applying the first 4 rules to produce the
configuration $[_0X[_A[_\$]_\$]_A]_0$.  We need the membrane labelled by $\$$ in
order to identify the end of the string.
\par Assume that we have a configuration of the form
$[_0X[_{x_1}[_{x_2}\ldots[_{x_h}[_\$]_\$]_{x_h}\ldots]_{x_2}]_{x_1}]_0$ after
some steps, where $h \geq 1$, and $Xx_1x_2\ldots x_h$ is a sentential form of
$G$, with $X\in N_1, x_i \in N_2 \cup T$.  Now we apply the rule $[_0X\ra
Y_i]_0$, for some $1 \leq i \leq n$.  We have two cases according to the values
of $i$.
\par\noindent {\it Case 1}: $1 \leq i \leq k$.  In this case, we are simulating a
matrix of type 2, i.e., $m_i : (X \ra Y, A \ra x)$.  By using the rule
$Y_i[_y]_y \ra [_yY_i]_y$, we move $Y_i$ deeper inside the nested membranes.  If
there is no membrane labelled by $A$  in the current configuration,
then we use the rule $Y_i[_\$]_\$ \ra [_\#\#]_\#$.  The symbol $\#$ generates an
infinite computation by means of the rule $[_\#\# \ra \#]_\#$.  If some membrane
labelled $A$ is present in the current configuration, then we have two cases.
\par \noindent {\it Case a)}: $|x|=1$, i.e., $x=a \in N_2 \cup T$.\\
In this case, we use the rule $Y_i[_A]_A \ra [_aY^\p_i]_a$.  The above rule
changes the object $Y_i$ into $Y^\p_i$ and also changes the label $A$ into $a$.
Now the object $Y^\p_i$ travel towards the skin membrane.  Once it reaches the
skin, it becomes $Y$.
\par \noindent {\it Case b)}: $|x|=2$, i.e., $x=a_1a_2 \in N_2 \cup T$.\\
In this case, we use the rule $Y_i[_A]_A \ra [_{a_1}Y_{i,a_2}]_{a_1}$.  Here the
object $Y_i$ changes to $Y_{i,{a_2}}$ and the label $A$ changes to $a_1$.
Further we use the rule $Y_{i,B}[_C]_C \ra [_BY_{i,C}]_B$ to move the object
$Y_{i,B}$ towards the membrane labelled~$\$$.  While moving the objects
$Y_{i,B}$, we change the labels of the membrane remembering the previous label
in their second component.  Once we got the object $Y_{i,\$}$ in the innermost
membrane, we use it to create a membrane labelled~$\$$ containing the object
$Y^\p_i$.  After this we move $Y^\p_i$ towards the skin membrane.  It will
become $Y$ once it reaches the skin membrane.
\par \noindent {\it Case 2}: $k+1 \leq i \leq n$.  That is we are simulating a matrix
of type 3 ($m_i: (X \ra Y, A \ra \#)$).  In this case, we use the object $Y_i$ to
check for the presence of a membrane labelled $A$ in the current configuration.
If there exists a membrane labelled $A$, the object $Y_i$ is moved inside by the
rule $Y_i[_A]_A \ra [_\#\#]_\#$.  This will lead to an infinite computation that
yields no result.  Otherwise, the object $Y_i$ becomes $Y^\p_i$ after reaching
the innermost membrane labelled by $\$$.  Now we move the object $Y^\p_i$
towards the skin membrane where it changes to $Y$.
\par Finally, we erase the symbol $Z$ in the skin membrane, once it was
introduced and the computation halts.  Now by applying rules as in the previous
theorem, we send out the objects in the right order.\qed
\section{Conclusion}
This paper explores the idea of defining membrane systems that are able to build
up a membrane
structure that encodes some meaningful information  proposed by \cite{lang-AM}.
We investigated the computational power
of P systems with membrane creation and dissolution rules operating according to
the external and traces mode.
Also we proved the universality of P systems with membrane creation alone, but
we allow the label
changing feature for {\em in} type rules.  At the moment, we are unable to
characterize the power
of P systems with active membranes equipped with membrane creation alone.

\bigskip

{\bf Acknowledgements.} M. Gheorghe and R. Rama are grateful to the Royal Academy
of Engineering which through a grant supporting {\em  Exchanges with India and
China} (2010), partially funded this research.~MG has been also
supported by CNCSIS grant no. 643/2009, {\em An integrated evolutionary approach
to formal modelling and testing}.
The authors would like to thank the anonymous reviewers for the comments made on an early version of this paper.

\bibliographystyle{mecbic}

\end{document}